\documentstyle[amsfonts,aps,eqsecnum]{revtex}
\begin{document}
\twocolumn[\hsize\textwidth\columnwidth\hsize\csname@twocolumnfalse%
\endcsname
\draft

\title{
Multiphoton Coincidence Spectroscopy
}

\author{L.\ Horvath, B.\ C.\ Sanders and B.\ F.\ Wielinga}
\address{
Department of Physics, Macquarie University \\
Sydney, New South Wales 2109, Australia \\}
\date{\today}
\maketitle

\begin{abstract}
We extend the analysis of photon coincidence spectroscopy beyond
bichromatic excitation and two-photon coincidence detection to include
multichromatic excitation and multiphoton coincidence detection.
Trichromatic excitation and three-photon coincidence spectroscopy are
studied in detail, and we identify an observable 
signature of a triple resonance in an atom-cavity system.

\end{abstract}


]

\narrowtext

\section{Introduction}
\label{sec:introduction}

Cavity quantum electrodynamics (CQED)\cite{Berman94}
in the optical domain is rapidly progressing:
advances in atom cooling methods, as well as improved
optical cavities which allow large dipole coupling
strengths, are leading experiments into new frontiers of research.
Single-atom experiments are now possible\cite{Hood98,Mabuchi98}, 
and trapping of atoms in optical cavities should soon be feasible\cite{Wong97}.
Quantum effects and quantitative testing of theoretical models for
the CQED system can be performed better than ever
and perhaps directed to certain applications
such as quantum logic gates\cite{Turchette95}.
Exciting developments are also taking place in the microwave
domain\cite{Brune96,Haroche97,Maitre97},
but here we are concerned with photon coincidence measurements,
which are performed only in the optical domain.

The method of photon coincidence spectroscopy (PCS) has been introduced as
a means to study the spectrum of the combined atom-cavity
system\cite{CLEO95,Carmichael96,Sanders97,Horvath99},
but this method was restricted to probing only the first couplet
of the nonlinear regime of the Jaynes-Cummings (JC) spectrum 
(see Fig.\ref{fig:ladder}).
Here we generalize the method of photon coincidence spectroscopy to
show how probing of higher levels of the spectrum can be performed,
and we show that three-photon coincidence spectroscopy (3PCS) could
yield a signature of the second couplet in
the nonlinear regime of the JC spectrum\cite{Jaynes63},
thus enabling direct, unambiguous probing of the quantum
features of a single atom in an optical cavity.

Photon coincidence spectroscopy is necessary to probe quantum features
of the atomic CQED system because practical difficulties limit
the efficacy of other techniques.
The major difficulties in the optical regime include the
width of the atomic beam traversing the cavity,
motion of the atom through the cavity,
and the interruption of the Rabi oscillation numerous times
during passage through the cavity as well as a fluctuating atomic number.
However, for sufficiently slow moving atoms\cite{Carmichael96},
the atoms can be regarded as being essentially stationary,
and the motion and spread of the atoms are responsible, then,
for an inhomogeneous broadening of the spectral peaks.
Furthermore, at low densities, single-atom effects dominate over
multi-atom process, and the JC model provides an excellent 
description.
Photon coincidence spectroscopy was then devised as a way
to probe the interesting quantum features in the presence
of unavoidable inhomogeneous broadening.

Until now, only two-photon coincidence spectroscopy (2PCS) has been studied,
both as quantum trajectory simulations\cite{Carmichael96} and
analytic, continued-fraction methods\cite{Sanders97,Horvath99}.
However, 2PCS is only useful for probing the second couplet
of the JC ladder, or, equivalently, the first couplet of the nonlinear regime 
(see Fig.\ref{fig:ladder}).
Our aim here is to propose multi-photon, or $N$-photon coincidence 
spectroscopy (NPCS) as 
a method for probing higher-level states and to show explicitly
how 3PCS would work and its feasibility.

\section{Dynamics}
\label{sec:dynamics}

The JC Hamiltonian for the atom-cavity system
\begin{equation}
\label{eqn:Hamiltonian}
\hat H(g) = \hbar \omega \hat{\cal N} + i \hbar g \hat{\cal A} , 
\end{equation}
where~$\omega$ is both the
atomic transition frequency and cavity resonance frequency,
$g$ is the dipole coupling strength,
\begin{equation}
\qquad \hat{\cal N} = \hat \sigma_3+\hat a^{\dag} \hat a + 1/2 
\end{equation}
is the ``excitation number'' operator, and
\begin{equation}
\hat{\cal  A} = \hat a^{\dag} \hat\sigma_- - \hat a \hat\sigma_+ =
-\hat{\cal A}^{\dag} ,
\end{equation}
describes the isolated atom-cavity system.
The coupling strength~$g$ is not actually a fixed constant
in the calculations: instead a distribution~$P(g)$ is constructed
which accounts for the variability of atomic position within the 
cavity mode and results in inhomogeneous broadening
\cite{Carmichael96,Sanders97}.
The master equation 
\begin{eqnarray}
\label{eqn:master}
\dot{\rho} &=& [ \hat H(g),\rho ] / i\hbar
	+ \left[ {\cal E}(t) \hat\sigma_+ - {\cal E}^*(t) \hat\sigma_-,\rho
	\right]
\nonumber \\
	&&+ (\gamma_I/2) (2\hat\sigma_- \rho \hat\sigma_+ - \hat\sigma_+ 
	\hat\sigma_- \rho - \rho \hat\sigma_+ \hat\sigma_- ) \nonumber \\
	&&+ \kappa (2\hat a\rho \hat a^{\dag} - \hat a^{\dag}\hat a\rho -
	\rho \hat a^{\dag} \hat a),
\end{eqnarray}
incorporates both the driving term~${\cal E}(t)$ and losses through
the cavity mirror and by fluorescence.
The solution~$\rho$ is $g$-dependent, and the final result
for the density matrix is
\begin{equation}
\rho = \int P(g) \rho(g) dg
\end{equation}
where~$\rho(g)$ represents the solution of
the master equation~(\ref{eqn:master}) for fixed~$g$.

In order to perform NPCS,
which is designed to probe the $N^{\rm th}$ couplet,
the driving field must be $N$-chromatic; {\em i.e.}
\begin{equation} 
{\mathcal E}(t) = \sum_{m=1}^N {\mathcal E}_m e^{-i \omega_m t } 
\end{equation} 
for~$\{{\cal E}_m\}$ a set of~$N$ constants, which are 
assumed to be real without loss of generality.
Moreover, the amplitudes~${\cal E}_m$ are sufficiently large to 
ensure significant 
occupation of the excited states but not so large that Stark 
shifts or occupation of the higher order ($n>N$) states is significant.
By judicious choice of each frequency~$\omega_m$, selective excitation to
the~$N^{\rm th}$ couplet is possible. 

In the rotating frame~$\omega_1 = 0$,
and we define detunings~$\delta_m \equiv \omega_m - \omega$. 

The randomness of the coupling strength~$\{g\}$ is responsible for
the inhomogeneous broadening depicted in Fig.~\ref{fig:ladder}, 
where~$g_{\rm max}$ is the coupling strength between the atom and
cavity mode at an antinode of the longitudinal axis. 
As the probability for the atom-field coupling strength being~$g_{\rm max}$
is quite small, we choose instead to selectively excite the
coupled atom--cavity system for some 
other~$g=g_f < g_{\rm max}$\footnote{The multiphoton spectroscopic
signal quality will depend on the choices of~$g_f$; for our simulation,
we choose a value which provides a good signal but not necessarily
the optimal signal.}.
Thus, we fix~$\omega_1=\omega+g_f$ as shown in
Fig.~\ref{fig:ladder}, and the amplitude of this field is, 
of course,~${\cal E}_1$.
More conveniently, the normalized detunings are defined by
\begin{equation}
\label{eqn:delta}
\tilde\delta_m \equiv \delta_m/g_f.
\end{equation}
We also define a normalised coupling strength as~$\tilde g=g/g_f$.
The master equation~(\ref{eqn:master}) can be expressed instead in terms
of the Liouville superoperator as 
\begin{eqnarray} 
\label{L(g,t)} 
{\mathcal L}(g,t) \rho(g,t)  
	&=& {\mathcal Q}(g) \rho(g,t)  
		+ \sum_{m=2}^N \Big( {\mathcal E}_m e^{-i 
	(\delta_m-g_f) t}  
		\Sigma_+ \nonumber	\\	&&
		- {\mathcal E}^*_m  e^{i (\delta_m-g_f) t}  
		\Sigma_- \Big) \rho(g,t)  
\end{eqnarray} 
where~$\Sigma_{\pm}\rho \equiv [ \sigma_{\pm},\rho ]$,
and~${\mathcal Q}(g)$ includes conservative and  dissipative superoperators.
The Bloch function method is applied by expanding
\begin{equation} 
\label{eqn:blocksol} 
\rho(g,t) 
	=\sum_{ \vec{k} \in {\Bbb Z}^{N-1} } 
		\rho_{\vec{k}}(g,t) e^{ - i \vec{k} \cdot 
	(\vec{\delta}-g_f \vec{1}) t }  
\end{equation}
for~${\Bbb Z}^{N-1}$ the set of all length $N-1$ vectors with integer values
and~$\vec{1}$ the vector with unity as every component.
Transient effects can be neglected;
hence, $\lim_{t \rightarrow \infty} \dot{\rho}_{\vec{k}}(g,t) \rightarrow 0$.
Thus, eq.~(\ref{L(g,t)}) reduces to 
\begin{eqnarray} 
\label{rho.equation} 
\sum_{m=2}^N {\mathcal E}_m \left(  \Sigma_+
	\rho_{\vec{k}-\vec{I}_{m-1}}(g)
	- \Sigma_- \rho_{\vec{k}-\vec{I}_{m+1}}(g) \right) 
		\nonumber	\\
	+\left[ i \vec{k} \cdot (\vec{\delta}-g_f\vec{1})  
	+ {\mathcal Q}(g) \right] \rho_{\vec{k}}(g) = 0  
\end{eqnarray}
where time dependence is ignored as~$t\rightarrow\infty$, and
$\vec{I}_m$ is a vector with all elements being~$0$ 
except the $m^{\rm th}$ element which is one.
Writing the superoperators~${\mathcal Q}(g)$ and~$\Sigma_{\pm}$ 
as matrices and~$\rho(g)$ as a vector,
eq.~(\ref{rho.equation}) represents infinitely many coupled linear equations.
In order to reduce the number of equations to a finite number, 
we introduce a positive integer~$q$ and establish the approximation
\[ \rho_{\vec{k}}(g) = 0 \;
	\forall \vec{k} \; {\rm satisfying}
	\sum_{i=1}^{N-1}|\vec{k}_i| > q, \]
which is valid for sufficiently small~$\{ {\cal E}_k \}$.  
We perform our expansion in the dressed state basis
$\{ |n)_{\pm} | n \in \{ 0 \} \cup {\Bbb Z}^+ \}$ which satisfies
\[ \hat{H}|n)_{\pm} = \hbar [ n\omega \pm \sqrt{n}g ] |n)_{\pm}, \;
\hat{H}|0) = 0 . \]
This spectrum of~$\hat{H}$ eigenvalues is shown in Fig.~\ref{fig:ladder}.
For NPCS, we truncate beyond the $N+1$ couplet. 
Hence, each coefficient~$\rho_{\vec{k}}(g)$ is of length~~$(1+2(N+1))^2$.
Setting~$q=1$ leaves~$5+2(N-3)$ matrix equations,
each square matrix of dimension~$(1+2(N+1))\times(1+2(N+1))$.

We are particularly interested in $\rho_{\vec{k}=\vec{0}}(g)$,
which is the `dc', or non-oscillating, component of the Bloch expansion.
In an experiment, the $N$-chromatic field would drive the atom-cavity system,
and $N$-quanta resonances would yield $N$-photon decays over
a timescale shorter than the cavity lifetime\cite{Horvath99}.
To a good approximation,
the $N$-photon coincidence rate is proportional to
$\langle a^{\dag \,N} a^{N} \rangle$,
and we evaluate this mean with respect to the density matrix 
component~$\rho_{\vec{k}=\vec{0}}(g)$. 
The elements of the density matrix are designated 
\begin{eqnarray}
\label{matrixelements}
\rho_{00}(g)&\equiv& (0|\rho_{\vec{k}=\vec{0}}(g)|0), 
\rho_{nn^\prime}^{\,\varepsilon\varepsilon^\prime}(g)\equiv
\,_{\varepsilon}(n|\rho_{\vec{k}=\vec{0}}(g)|n^\prime)_{\varepsilon^\prime},
\nonumber \\
\rho_{0n}^{\; \; \varepsilon}(g)&\equiv& (0|\rho_{\vec{k}=\vec{0}}(g)
|n)_{\varepsilon}\equiv\rho_{n0}^{\,\varepsilon}(g)^{*}.
\end{eqnarray}
The resonance would be observed in practice by fixing~$\tilde\delta_i$
for~$i=2,3,\ldots,N-1$, and varying~$\tilde\delta_N$.
The $N^{\rm th}$ couplet is then observable experimentally as
an increase in the $N$-photon coincidence rate as a
function of $\tilde\delta_N$; i.e. 
as~$\left\langle a^{\dag\,N} a^{N} \right\rangle$
{\em vs}~$\tilde\delta_{N}$.

However, excited state resonances are complicated by the existence
of off-resonant excitations which result in spurious
$N$-photon decays.
To study these resonant and off-resonant effects in greater detail,
we calculate multiphoton peak heights at various values
of $\tilde\delta_N$ using the non-Hermitian Hamiltonian formalism.
The master equation~(\ref{eqn:master}) incorporates a non-unitary
evolution which can be treated as a combination of `loss'
terms and `jump' terms,
in the sense of quantum trajectories\cite{Carmichael96}.
The non-Hermitian Hamiltonian is\cite{Horvath99}
\begin{eqnarray}
\label{Heff}
H_{\rm eff}(g)
	&=& \left( \omega - \omega_1 \right) \hat{\cal N} 
	+ i g \hat{\cal A}
	+ i \sum_{m=1}^{N}{\cal E}_m  
	(e^{-i (\delta_m-g_f)t} \sigma_+ \nonumber \\
	&& - e^{i (\delta_m-g_f)t}\sigma_- )
	- \kappa a^{\dagger} a - (\gamma_I/2) \sigma_+ \sigma_-.
\end{eqnarray}
We expand the Hamiltonian as a matrix in the truncated
dressed-state basis, where coefficients of states in the~$(N+1)^{\rm th}$
couplet and higher are ignored:
the time-dependent state is approximated by
\begin{equation}
\label{statevector}
|\psi(g,t)\rangle \approx c_0(g,t) |0)
	+ \sum_{ n=1 }^3 \sum_{ \varepsilon = \pm } c_{n}^{\varepsilon}(g,t)
	|n)_{n\varepsilon} 
\end{equation}
with~$\left\langle \psi(g,t)| \psi(g,t) \right\rangle \leq 1$~$\forall$
~$t\geq0$ and~$c_0(t=0)=1$ and~$c_n^{\varepsilon}(t=0)=0$.
We work in the rotating-wave approximation,
and the~$2N+1$ coefficients~$\{ c_0, \{ c_{n}^{\varepsilon} \} \}$ 
can be written as a vector~$\vec{c}(g,t)$.
The matrix differential 
equation is~$\stackrel{.}{\vec{c}}\!(g,t) = M(g,t) \vec{c}\,(g,t)$.
We can write the matrix as 
\begin{equation}
M(g,t) = \sum_{\ell=0}^{L} M_{\ell} e^{-i\Omega_{\ell} t},
\end{equation}
where~$\vec{\Omega}$ is a vector of unequally spaced discrete frequencies,
ordered from~$\Omega_0=0$ to ever-increasing values of frequency.
That is, $\Omega_i > \Omega_j$ for $i > j$.
In solving the equation, we ignore terms~$\Omega_{\ell>L}$
where~$L$ is a cut-off parameter.
Physically this corresponds to retaining terms responsible for
Stark shifts in levels up to order~$L$.
Truncating this expansion is valid because we assume that
the amplitudes of the driving fields~$\left\{ {\cal E}_m \right \}$ are small.
 
This calculation of~$\{ c_0, \{ c_{n}^{\varepsilon} \} \}$ allows us
to approximate the $N$-photon count rate
(NPCR)~$\left\langle a^{\dag N} a^{N} \right\rangle$
for given values of~$\tilde\delta_N$ where peaks are observed in the 
full simulation following from eq.~(\ref{rho.equation}). We use this
analysis to verify the validity of the computer simulation, applied to
the special case~$P(g)=\delta(g-g_f)$, and to observe the importance of 
the Stark effect on the peak heights. 
As we show below for 3PCS, we have excellent agreement between numerical
simulations and this semianalytic approach using the non-Hermitian Hamiltonian
formalism.

\section{Three-Photon Coincidence Spectroscopy}
\label{sec:3PCS}

For large~$N$, significant computer time and memory is required to solve 
the equations, but 3PCS, corresponding to $N=3$,
is readily solved.
To probe the third couplet of the JC ladder, a trichromatic driving field
is employed. Ideally a photon of 
frequency~$\omega_1= \omega+g_f$ 
induces the transition~$|0) \rightarrow |1)_+$, followed by 
another photon of frequency~$\omega_2 = \omega + (\sqrt{2}-1)g_f$ 
which takes the excitation from the~$|1)_+$ to the~$|2)_+$ state, 
and, finally, a third photon of frequency~$\omega_3$ scans the
system over a range of frequencies including the~$|2)_+\longleftrightarrow 
|3)_\pm $ transitions as shown in 
Fig.~\ref{fig:ladder}.

By setting~$N=3$,
eq.~(\ref{rho.equation})
reduces to
\begin{eqnarray} 
\label{rho.equation3} 
0 &=& \left[ i \left( k_2 \delta_2 + k_3 \delta_3 \right)  
	+ {\mathcal Q}(g) \right] \rho_{k_2, k_3}(g) 
	+ {\mathcal E}_2 \Sigma_+ \rho_{k_2-1,k_3}(g) \nonumber \\ 
	&&- {\mathcal E}_2 \Sigma_- \rho_{k_2+1,k_3}(g) 
	+ {\mathcal E}_3 \Sigma_+ \rho_{k_2,k_3-1}(g) 
		\nonumber	\\	&&
	- {\mathcal E}_3 \Sigma_- \rho_{k_2,k_3+1}(g). 
\end{eqnarray} 
Applying the approximation~$\rho_{k_2,k_3}(g) = 0$ 
for~$|k_2| +|k_3| > q$ reduces the number of coupled matrix equations to
$n(q) = 2q^2 + 2q + 1$, where coefficients corresponding to
the fifth couplet and higher are ignored.
Hence, there are~$81$ scalar coefficients of~$\rho_{k_2,k_3}(g)$. 
In order to reduce computing time, 
we set~$q=1$ thereby yielding 
2025 simultaneous equations. 

The signature of genuine three-photon decay can be obtained by measuring
the three-photon count rate
(3PCR),~$\langle \hat{a}^{\dag 3}\hat{a}^3 \rangle$, 
{\em vs}~$\tilde\delta_3$ to observe three-photon resonance peaks.
The density matrix elements and the 3PCR are shown in Fig.~\ref{ninegraphs}
for a range of~$\tilde g$. 
In Fig.~\ref{ninegraphs}(a) we observe two important features for~$\rho_{00}$.
Firstly, there are two valleys located at~$\tilde\delta_3=\pm \tilde g$ due to 
the vacuum Rabi splitting effect.  
Secondly, there are  two valleys located at~$\tilde g=1$ and~$\tilde g=\sqrt{2}-1$,
independent of~$\tilde\delta_3$. The former valley due to off-resonant
three-photon excitation to the third couplet from the ground state via
the pathway~$\omega_1\rightarrow\omega_2\rightarrow\omega_2$ 
(a photon with frequency~$\omega_1$ resonantly excites~$|0)$ to~$|1)_+$,
followed by resonant excitation by a photon with frequency~$\omega_2$
to~$|2)_+$ and finally off-resonant excitation by a photon with 
frequency~$\omega_2$ to~$|3)_+$). The energy of the state~$|3)_+$ 
is~$\hbar(3\omega+(2\sqrt{2}-1)g_f)$ for~$\tilde g=1$. The sum of
the energies in the three photons producing 
the~$\omega_1\rightarrow\omega_2\rightarrow\omega_2$ off-resonant
excitation pathway is~$\hbar(3\omega+\sqrt{3}g_f)$. 
The detuning is thus~$(2\sqrt{2}-1-\sqrt{3})g_f\doteq 0.1g_f$, which
is quite small, thus ensuring the significant depletion observed 
in~Fig.~\ref{ninegraphs}(a). 
In Fig.~\ref{ninegraphs}(b), where~$\rho_{\,3\,3}^{++}$ is plotted, a
ridge is observed at~$\tilde g=1$, and the off-resonant excitation 
to~$\rho_{33}^{++}$ is thus clear.

The second valley in~Fig.~\ref{ninegraphs}(a) occurs at~$\tilde g=\sqrt{2}-1$.
This depletion from the ground state arises due to resonant excitation
from the ground state~$|0)$ to the excited state~$|1)_+$ via
absorption of a photon of frequency~$\omega_2$ which is fixed
(independent of~$\tilde\delta_3$). In contradistinction to the presence of
a ridge at~$\tilde g=1$ in~Fig.~\ref{ninegraphs}(b), a ridge is {\em not}
observed in Fig.~\ref{ninegraphs}(b) at~$\tilde g=\sqrt{2}-1$ because
excitation to the second couplet is off-resonant. On the other hand 
excitation to the second couplet is resonant for~$\tilde g=1$ and so 
the ridge is visible for~$\tilde g=1$ in Fig.~\ref{ninegraphs}(b).

The two valleys in~Fig.~\ref{ninegraphs}(a) at~$\tilde g=1$ and~$\tilde g=\sqrt{2}-1$
have different depths. As each valley is induced by resonant excitation
from~$|0)$ to~$|1)_+$, the depletion of~$\rho_{00}$ can be calculated
from a two-state approximation~\cite{Sanders97,Tian92}
\begin{eqnarray}
\label{approx}
\rho_{00}\doteq 1
-\frac{{\cal E}^2_m}{\frac{1}{2}(\kappa+\gamma_I/2)^2+2 {\cal E}^2_m},
m=1,2. \nonumber
\end{eqnarray}
For~$\tilde g=1$,~$m=1$ and~${\cal E}_1=1/\sqrt{2}$ 
producing~$\rho_{00}\doteq 13/17$ as observed in Fig.~\ref{ninegraphs}(a).
Similarly, for~$\tilde g=\sqrt{2}-1$,~$m=2$ and~${\cal E}_2=\sqrt{2}$ 
producing~$\rho_{00}\doteq 25/41$ which again matches 
Fig.~\ref{ninegraphs}(a).

A subtle feature of Fig.~\ref{ninegraphs}(b) is the presence of two dips
along the ridge at~$\tilde g=1$. One dip occurs at~$\tilde\delta_3=-1$.
This dip is due to competition between two excitation pathways. One path 
involves
resonant excitation~via the~$\omega_1$ photon to~$|1)_+$, and the
second path is resonant excitation to~$|1)_-$~via the~$\omega_3$ photon. 
Excitation to~$|1)_-$ diminishes the probability of excitation 
to~$|1)_+$ which is necessary for 
the~$\omega_1\rightarrow\omega_2\rightarrow\omega_2$ off-resonant 
excitation to~$|3)_+$, hence the dip in~$\rho_{\,3\,3}^{++}$.

The second dip occurs at~$\tilde\delta_3=-(\sqrt{2}+1)$ which also
occurs due to competition between paths; however, this dip is less 
noticeable. Both paths experience resonant excitation to~$|1)_+$, but
the~$\omega_3$ photon excites to~$|2)_-$ in competition with 
the~$\omega_2\rightarrow\omega_2$ two-photon subsequent excitation to~$|3)_+$.
As the competition occurs for electrons in the~$|1)_+$ state, instead of
for the~$|0)$ state in the case of the other dip, the competition, and hence
the dip, is less significant.

Two prominent off-resonant peaks are centred 
about~$(\tilde\delta_3,\tilde g)=(1,1)$ and the other 
at~$(\tilde\delta_3,\tilde g)=((\sqrt{2}-1)^2,\sqrt{2}-1)$. 
The former peak is due to 
cooperative excitation pathways via the three-photon 
excitations~$\omega_1\rightarrow\omega_2\rightarrow\omega_2$ 
and~$\omega_3\rightarrow\omega_2\rightarrow\omega_2$. The equality 
between~$\omega_3$ and~$\omega_1$ is responsible for this 
cooperative effect. The second off-resonant peak is due to 
cooperation between the 
pathways~$\omega_2\rightarrow\omega_2\rightarrow\omega_2$
and ~$\omega_2\rightarrow\omega_3\rightarrow\omega_3$. The ridge
at~$\tilde g=\sqrt{2}-1$ due to 
the~$\omega_2\rightarrow\omega_2\rightarrow\omega_2$ pathway is, however,
negligible due to off-resonant excitation to the second couplet as well as
to the third couplet. 

There is a prominent peak 
at~$(\tilde\delta_3,\tilde g)=(\sqrt{3}-\sqrt{2},1)$ in 
Fig.~\ref{ninegraphs}(a), corresponding  to cooperative excitation 
pathways~$\omega_1\rightarrow\omega_2\rightarrow\omega_3$ 
and~$\omega_1\rightarrow\omega_3\rightarrow\omega_2$ for resonant
excitation to~$|3)_+$. 
The off-resonant excitation 
pathway~$\omega_1\rightarrow\omega_2\rightarrow\omega_2$ also occurs.

Finally, we observe a peak for~$\tilde g\leq 0.1$ and centred 
at~$\tilde\delta_3=0$. This peak is very large near~$\tilde g=0$ 
(not shown) and corresponds to very small splitting of the
couplets. Consequently, the system behaves much like a decoupled
atom and cavity and acts as a resonator for~$\tilde\delta_3=0$.
The 3PCR, proportional 
to~$\left\langle a^{\dag\,3}a^3 \right \rangle$, is plotted in 
Fig.~\ref{ninegraphs}(c).
The similarity between Figs.~\ref{ninegraphs}(b) and~\ref{ninegraphs}(c) is
evidence that occupation of~$|3)_+$ is a good indicator of the 3PCR.
However, occupation of~$|3)_-$ also contributes to the 3PCR.
The peak at~$\tilde\delta_3=-(\sqrt{3}+\sqrt{2})$ in Fig.~\ref{ninegraphs}(c)
is due to the excitation 
pathway~$\omega_1\rightarrow\omega_2\rightarrow\omega_3$ which resonantly
excites to~$|3)_-$.
The peak at~$\tilde\delta_3=-(\sqrt{3}+\sqrt{2})$ is somewhat diminished, 
however, by the off-resonant~$\omega_1\rightarrow\omega_2\rightarrow\omega_2$
excitation pathway. The peak at~$\tilde g\leq 0.1$ is significantly larger in
Fig.~\ref{ninegraphs}(c) than in Fig.~\ref{ninegraphs}(b) due to
contributions from off-resonant excitation to both~$|3)_-$ and~$|3)_+$.

The peaks, valleys and ridges in Figs.~\ref{ninegraphs} have been explained
in terms of excitation pathways. We have introduced each of these pathways
by studying the system intuitively, but a verification is possible using
the effective non-Hermitian Hamiltonian~(\ref{Heff}) and the approximate
time-dependent unnormalized state (\ref{statevector}). In Table~\ref{table1}
we presented analytical estimates of the 
3PCR~$\left \langle a^{\dag\,3}a^3 \right \rangle_{\rm est}$,
using the non-Hermitian Hamiltonian, and compare to the 
3PCR,~$\left \langle a^{\dag\,3}a^3 \right \rangle$, observed in
Fig.~\ref{ninegraphs}(c). Although the estimates can vary from the 
observed 3PCR by up to a factor of~$2.5$, the agreement is excellent
considering that `jump' terms have been ignored which significantly
decrease the peak height. An exception is the dip 
at~$(\tilde\delta_3,\tilde g) =(-1,1)$ where the `jump' terms are 
responsible for an increase in peak height.
Table~\ref{table1} provides confirmation of the pathways suggested as
being responsible for features in Figs.~\ref{ninegraphs}. In each case
the cut-off parameter is~$L=1$, except 
for~$\tilde\delta_3=-(\sqrt{3}+\sqrt{2})$
where~$L=2$ is due to two pathways: one resonantly exciting to~$|3)_-$ 
and the other due to off-resonant excitation to~$|3)_+$.

In Figs.~\ref{smeared}(a) and~\ref{smeared}(b) we present the 3PCR 
{\em vs}~$\tilde\delta_3$, averaged over~$P(g)$, assuming a TEM$_{00}$
mode in the cavity and a mask for the atomic beam. The expression 
for~$P(g)$ is complicated and is provided in the appendix of 
Ref.~\cite{Sanders97}.
The average 3PCR is thus
\begin{equation}
\label{eq:average}
\overline{\left\langle a^{\dag\,3}a^3 \right\rangle}=\int Tr\left(
\rho(g)a^{\dag\,3}a^3\right)P(g)dg.
\end{equation}
Fig.~\ref{smeared}(a) corresponds to a strong coupling of~$g_f/\kappa=63$.
This strong coupling ensures that fine features are not destroyed by 
inhomogeneous broadening. By comparison, the case~$g_f/\kappa=9$ is
presented in Fig.~\ref{smeared}(b), and the deleterious effects of
inhomogeneous broadening are evidently much stronger for the weaker-coupling
case.

In each graph the dotted line corresponds to the 3PCR {\em after} background
subtraction. To perform background subtraction, the experiment is performed 
four times, once with all three field on and scanned over~$\tilde\delta_3$.
The experiment is then repeated for the three cases 
(i)~${\cal E}_1={\cal E}_2=0$,
(ii)~${\cal E}_1\neq 0$ and~${\cal E}_2=0$ and
(iii)~${\cal E}_1=0$ and~${\cal E}_2\neq 0$. 
The contribution to the 3PCR is
then due solely to the excitations 
by (i)~$\omega_3$ photons, (ii) $\omega_1$ and $\omega_3$ photons, and
(iii) $\omega_2$ and $\omega_3$ photons. Case (i) is a subset of (ii) 
and (iii), 
hence must be subtracted
from (ii) and (iii). Using the notation of Ref.~\cite{Sanders97}, we denote
the spectrum after subtraction as

\begin{eqnarray}
\label{background}
\Delta^{(3)}(\tilde\delta_3)\equiv&&
\overline{\left\langle a^{\dag\,3}a^{3}\right\rangle}(\tilde\delta_3)-
\overline{\left\langle a^{\dag\,3}a^{3}\right\rangle}_{{\cal E}_1=0}
(\tilde\delta_3) \nonumber \\
 - &&
\overline{\left\langle a^{\dag\,3}a^3\right\rangle}_{{\cal E}_2=0}
(\tilde\delta_3) +
\overline{\left\langle a^{\dag\,3}a^3\right\rangle}_{{\cal E}_1=0,
{\cal E}_2=0}(\tilde\delta_3).
\end{eqnarray}

In Fig.~\ref{smeared}, the value of background subtraction is apparent. 
The 3PCR for the strong-coupling case~$g_f/\kappa=63$, depicted 
in~Fig.~\ref{smeared}(a), is improved by background subtraction. 
Broadening about~$\tilde\delta_3=0$ is reduced, and the peak 
at~$\tilde\delta=\sqrt{3}-\sqrt{2}$ is more evident after background
subtraction. Furthermore, the multiple peak structure near~$\tilde \delta_3=0$,
before background subtraction, consists of undesirable off-resonant
contributions. These are effectively removed by background subtraction,
and the dip in the 3PCR at~$\tilde\delta_3=1$ is evident. The reason for
this dip is that~$\omega_3=\omega_1$, and the two 
pathways~$\omega_1\rightarrow\omega_2\rightarrow\omega_2$ (which is 
responsible for the background 3PCR, independent of the value 
of~$\tilde\delta_3$) and~$\omega_3\rightarrow\omega_2\rightarrow\omega_2$
are complementary. Of course the most important peak occurs 
at~$\tilde\delta=-(\sqrt{3}+\sqrt{2})$, which is outside the inhomogeneous
broadening region. Observing this peak would not require the
time-consuming background subtraction methods necessary for discerning the
other peaks.
This peak at~$\tilde\delta_3=-(\sqrt{3}+\sqrt{2})$ exhibits the desired
``$\sqrt{3}$'' signature for excitation to the third couplet as well as
the ``$\sqrt{2}$'' signature arising from the excitation from the second
couplet.

One of the finer features in Fig.~\ref{smeared}(a) is the dip 
at~$\tilde\delta_3=-1$. This dip is due to the significant dip in 
Fig.~\ref{ninegraphs}(c) for~~$(\tilde\delta_3,\tilde g) =(-1,1)$. However,
the size of the dip is somewhat reduced due to a very small peak 
at~$(\tilde\delta_3,\tilde g)=(-1,\sqrt{2}-1)$ in Fig~\ref{ninegraphs}(c)
due to the off-resonant excitation 
pathway~$\omega_2\rightarrow\omega_3\rightarrow\omega_1$ to~$|3)_+$. 
Although this peak is quite small,~$P(g)$ is more highly weighted for
low~$g$. 

For Fig.~\ref{smeared}(a) the strong coupling case~$g_f/\kappa=63$ has been
adopted. The importance of strong coupling is that homogeneous broadening
due to  widths~$\gamma_I/2$ and~$\kappa$ are small, and contributions
due to off-resonant transitions are less significant. In previous
analyses of PCS, albeit for the two-photon case, the coupling
strength~$g_f/\kappa=9$ has been adopted. For 3PCS, such a coupling
is too small. In Fig.~\ref{smeared}(b) some of the more dramatic
features of Fig.~\ref{smeared}(a) are still discernible but degraded
to a level of near indistinguishability from the background. We 
observe the dips at~$\tilde\delta=\pm 1$ and the (very broad) peak 
at~$\tilde\delta=-(\sqrt{3}+\sqrt{2})$. However, experimental observation
of such features is unlikely, and a higher coupling strength is desirable.
A coupling strength of~$g_f/\kappa=63$ is not required, but a coupling
strength higher than~$g_f/\kappa=9$ is necessary.


\section{Conclusion}
\label{sec:conclusion}

The technique of photon coincidence spectroscopy (PCS) has been extended
from driving the atom-cavity coupled system by a bichromatic field and
measuring two-photon 
coincidences~\cite{CLEO95,Carmichael96,Sanders97,Horvath99}
to multichromatic driving fields and
multichromatic coincidences. Whereas two-photon coincidence spectroscopy (2PCS)
enables observation of the~$2\sqrt{2}g$ splitting of the spectrum,
associated with the second couplet, higher-order photon coincidence
spectroscopy allows direct probing of higher couplets in the JC ladder.
This technique thus offers a valuable tool for resolving higher-order
spectral phenomena in cavity quantum electrodynamics. Although the focus
here has been on an atomic beam and 
the spectrum associated with the
JC ladder, this scheme can be
adapted to studying spectra of other cavity QED systems with a 
discrete spectrum and significant inhomogeneous broadening.

We can also see that off-resonant phenomena are increasingly important
for probing higher-order couplets. The background subtraction scheme is
more intricate. Also resolving peaks for higher-order couplets requires
increasingly large coupling strengths. The required coupling strength
places a bound on the feasibility of $N$-photon coincidence spectroscopy,
and we see that a large coupling strength is required even for~$N=3$.

\section*{Acknowledgement}

We have benefited from valuable discussions with H.\ J.\ Carmichael, 
Z.\ Ficek and K.-P.\ Marzlin.
This research has been supported by a Macquarie University Research Grant
and by an Australian Research Council Small Grant.

\newpage
\begin{figure}
	\caption{
	A three-photon excitation scheme from the 
	ground state~$|0)$ to the first three couplets~$|n)_{\varepsilon}$
	($n\leq3$)
	of the dressed states with inhomogeneously broadened energy bands. 
	}
\label{fig:ladder}
\end{figure} 

\begin{figure}
\caption{	
	Plots of (a)~$\rho_{00}$, (b)~$\rho^{++}_{\,3\,3}$
	and (c) the~$\left\langle a^{\dag\,3}a^3\right\rangle$ {\em vs}  
	normalised scanning field
	frequency,~$\tilde{\delta}_3$, 
	and coupling strength,~$\tilde g$, 
	for~$\gamma_I/\kappa=1$,~$g_f/\kappa=63$,
	${\cal E}_1/\kappa=1/\sqrt{2}$ and 
	${\cal E}_2/\kappa={\cal E}_3/\kappa=\sqrt{2}$.
	}
\label{ninegraphs}
\end{figure}

\begin{figure}
\caption{Plots of~$\left\langle a^{\dag\,3}a^3\right\rangle$ with (solid)
	and without (dots) background subtraction 
	{\em vs}~$\tilde{\delta}_3/\pi$
	for~$\gamma_I/\kappa=1$,~${\cal E}_1/\kappa=1/\sqrt{2}$,
	~${\cal E}_2/\kappa={\cal E}_3/\kappa=\sqrt{2}$ 
	averaged over $P(g)$ for
	randomly-placed atom in the TEM$_{00}$ mode for (a)~$g_f/\kappa=63$ 
	and (b)~$g_f/\kappa=9$.}
\label{smeared}

\end{figure}

\begin{table}
\def\tstrut{\vrule height0pt depth5pt width0pt}
\begin{center}
\begin{tabular}{|c|c|c|c|} \hline\tstrut
$\tilde g$ & $\tilde\delta_3$ & 
$\left\langle a^{\dag\,3}a^3\right\rangle$ &
$\left\langle a^{\dag\,3}a^3\right\rangle_{\rm est}$\tstrut \\ \hline
$(\sqrt{2}-1)$ & $(\sqrt{2}-1)^2$ & $1.2\times 10^{-3}$ & $3.1\times 10^{-3}$\\ \hline
$1$ & $1$ & $1.7\times 10^{-3}$ &  $1.7\times 10^{-3}$\\ \hline
$1$ & $\sqrt{3}-\sqrt{2}$ & $1.5\times 10^{-3}$ & $2.6\times 10^{-3}$ \\ \hline
$1$ & $-1$ & $2.4\times 10^{-4}$ & $2.1\times 10^{-4}$\\ \hline
$1$ & $-(\sqrt{2}+1)$ & $3.3\times 10^{-4}$ &  $3.3\times 10^{-4}$  \\ \hline
$1$ & $-(\sqrt{3}+\sqrt{2})$ & $1.3\times 10^{-3}$ & $2.1\times 10^{-3}$\\ \hline
\end{tabular}
\end{center}
\caption{Three-photon count rate 
(3PCR)~$\left\langle a^{\dag\,3}a^3\right \rangle$
and approximate 3PCR~$\left\langle a^{\dag\,3}a^3\right \rangle_{\rm est}$
calculated via the non-Hermitian Hamiltonian formalism.} 
\label{table1}
\end{table}

\end{document}